\documentclass[superscriptaddress,secnumarabic,nobibnotes,aps,prd,showkeys,noshowpacs,onecolumn,12pt]{revtex4}
\usepackage{graphics}
\usepackage{graphicx}
\usepackage{color}
\usepackage{epsf}
\usepackage{bm}
\usepackage{amsmath,amssymb,amsfonts,amsthm}
\usepackage{latexsym}
\usepackage{enumerate}
\usepackage{hyperref}

\begin{document}

\title{Generalized Scale factor Duality Symmetry in Symmetric Teleparallel
Scalar-tensor FLRW Cosmology}
\author{Andronikos Paliathanasis}
\email{anpaliat@phys.uoa.gr}
\affiliation{Institute of Systems Science, Durban University of Technology, Durban 4000,
South Africa}
\affiliation{Departamento de Matem\'{a}ticas, Universidad Cat\'{o}lica del Norte, Avda.
Angamos 0610, Casilla 1280 Antofagasta, Chile}

\begin{abstract}
We review the Gasperini-Veneziano scale factor duality symmetry for the
dilaton field in scalar-tensor theory and its extension in teleparallelism.
Within the framework of symmetric teleparallel scalar-tensor theory, we
consider a spatially flat Friedmann--Lema\^{\i}tre--Robertson--Walker metric
cosmology. For the three possible connections, we write the corresponding
point-like Lagrangians for the gravitational field equations, and we
construct discrete transformations which generalize the Gasperini-Veneziano
scale factor duality symmetry. The discrete transformations depend on the
parameter which defines the coupling between the scalar field and the
nonmetricity scalar. The Gasperini-Veneziano duality symmetry is recovered
for a specific limit of this free parameter. Furthermore, we derive the
conservation laws for the classical field equations for these models, and we
present the origin of the discrete transformations. Finally, we discuss the
integrability properties of the model, and exact solutions are determined.
\end{abstract}

\maketitle

\section{Introduction}

\label{sec1}

A main characteristic of the two-dimensional conformal field theory is the
existence of duality symmetry. This transformation, which keeps the
invariant action principle, has important consequences in various aspects of
string theory and in modern cosmology \cite{tt1,tt2}. 

In this study, we refer to the T-duality symmetry introduced by Buscher \cite%
{Buscher1,Buscher2}. Specifically, when the "radius" $\mathcal{R}$~of the
geometry changes such that $\bar{\mathcal{R}}\rightarrow \mathcal{R}^{-1}$,
then for the two-dimensional $\sigma $-model, the field equations remain
invariant. The existence of this discrete symmetry is due to the appearance
of an isometry for the extended world-sheet. For more details, we refer the
reader to \cite{sd1,sd2,rvduality,rvd2}.

In \cite{gasp0}, Gasperini and Veneziano, by using the property of
T-duality, were able to construct a scale-factor duality transformation for
the dilaton Action Integral \cite{faraonibook} in the case of a spatially
flat Friedmann--Lema\^{\i}tre--Robertson--Walker metric \ (FLRW) geometry.
This discrete symmetry allows one to connect two different topological
spaces, and as a result, to connect different eras of the universe. The
duality transformation has been applied to studies of cosmological
observations \cite{gasp1}, as well as to explore the pre-big bang epoch of
the universe \cite{Gasp}. Recently, in \cite{g2023}, this property was used
to construct a family of nonsingular, nonperturbative pre-big bang
cosmological models. The duality transformation is due to the existence of
the $O\left( d,d\right) $ symmetry for the gravitational action of the
scalar-tensor model \cite{Shapere}.

Nowadays, the puzzle of the nature\ and the origin of the dark energy which
drives the current acceleration phase of the universe \cite{jo1,jo} has led
to the geometric modification of Einstein's gravitational theory by
introducing new invariants related to a more general connection. The Ricci
scalar $R$, the Torsion scalar $T$ and the Nonmetricity scalar $Q$, which
are determined by a general connection $\Gamma _{\mu \nu }^{\kappa }$, form
the so-called trinity of gravity \cite{tr1,tr2,tr3,tr4}. The Levi-Civita
components of the general connection lead to the Ricci scalar, the
antisymmetric components lead to the torsion scalar, while the nonmetricity
scalar is constructed by the remaining terms. In Einstein's gravity, the
connection is identical to the Levi-Civita connection; in teleparallelism,
the connection is antisymmetric; while in symmetric teleparallel theory, the
connection is symmetric and flat \cite{eisn1}.

These three scalars differ by boundary terms \cite{eisn1}, which means that
when they are used to define an Action Integral, the variation principle
leads to the same gravitational field equations. These are General
Relativity (GR), the Teleparallel Equivalent of GR (TEGR) \cite{hh1}, and
the Symmetric Teleparallel Equivalent of GR (STEGR) \cite{hh2}. However,
this equivalence is violated when additional degrees of freedom, that is,
scalar fields nonminimally coupled to the gravitational Lagrangian, are
introduced into the model. In \cite{pOdd}, the dilaton field, within the
framework of teleparallelism, was introduced, and the existence of the $%
O\left( d,d\right) $ symmetry was examined. It was found that a discrete
transformation exists which is a generalized Gasperini-Veneziano scale
factor duality transformation.

In this study, we focus on the dilaton field coupled to the nonmetricity
scalar $Q$, and we investigate the existence of duality transformations. For
the spatially flat FLRW background geometry, we consider the three possible
connections and determine the three different Lagrangian functions. We found
that discrete transformations exist, which generalize the
Gasperini-Veneziano scale factor duality symmetry. The transformations
depend on a constant which defines the coupling between the scalar field and
the nonmetricity scalar. There exists a limit for this parameter where the
generalized duality symmetry reduces to the Gasperini-Veneziano
transformation. Furthermore, we examined the existence of a continuous
transformation which serves as the origin of the discrete symmetry. Finally,
we focus on the analysis of the integrability properties of this
gravitational model. The structure of the paper is as follows.

In Section \ref{sec2}, we introduce the T-duality transformation for the nonlinear $\sigma$-mdodel. In Section \ref{sec2a}, we review the Gasperini-Veneziano scale
factor duality transformation for scalar-tensor cosmology and its extension
to teleparallelism. The symmetric teleparallel scalar-tensor model in a FLRW
cosmological background is introduced in Section \ref{sec3}.

Furthermore, in Section \ref{sec4}, we find that a generalized
Veneziano-Gasperini duality transformation exists in nonmetricity theories.
The existence of the generalized transformation is equivalent to the
existence of a conservation law for the classical field equations. In
Section \ref{sec5}, we determine the corresponding conservation laws for the
cosmological field equations, and we investigate their integrability
properties. With the use of canonical transformations, in Section \ref{sec5a}%
, we construct exact cosmological solutions within nonmetricity theories.
Finally, in Section \ref{sec6}, we summarize our results and draw our
conclusions.

\section{Duality symmetry}

\label{sec2}

In a $D-$dimensional Riemannian manifold with Lorentzian signature, we
define the bosonic nonlinear $\sigma -$model with Action Integral%
\begin{equation}
S=\frac{1}{4\pi a_{0}}\int d^{2}\xi \left( \sqrt{\gamma }\gamma ^{\mu \nu
}g_{ab}\partial _{\mu }x^{\alpha }\partial _{\nu }x^{b}+i\varepsilon ^{\mu
\nu }h_{ab}\partial _{\mu }x^{\alpha }\partial _{\nu }x^{b}+a_{0}\sqrt{%
\gamma }R^{\left( 2\right) }\phi \left( x^{\gamma }\right) \right) ,
\label{dd.01}
\end{equation}%
in which $\gamma _{ab}$ is a two-dimensional metric which describes the
world sheet with Ricci scalar $R^{\left( 2\right) }$, $g_{ab}$ is the target
metric, $h_{ab}$ is the torsion, $\phi $ is the dilaton field; coefficient
constant $a_{0}$ is the inverse of the string tension.

The Action Integral (\ref{dd.01}) can be constructed by the $1+D$%
-dimensional nonlinear $\sigma $-model%
\begin{eqnarray}
S_{\left( 1+d\right) } &=&\frac{1}{4\pi a_{0}}\int \left( \sqrt{\gamma }%
\gamma ^{\mu \nu }\left( g_{00}V_{\mu }V_{\nu }+2g_{0a}V_{\mu }\partial
_{\nu }x^{\alpha }+g_{ab}\partial _{\mu }x^{\alpha }\partial _{\nu
}x^{b}\right) \right)  \notag \\
&&+\frac{1}{4\pi a_{0}}\int d^{2}\xi \left( i\varepsilon ^{\mu \nu }\left(
2h_{0a}V_{\mu }\partial _{\mu }x^{a}+h_{ab}\partial _{\mu }x^{\alpha
}\partial _{\nu }x^{b}\right) +2\tilde{\theta}i\varepsilon ^{\mu \nu
}\partial _{\mu }V_{\nu }+a_{0}\sqrt{\gamma }R^{\left( 2\right) }\phi \left(
x^{\gamma }\right) \right)
\end{eqnarray}%
where $V_{\mu }$ is an 1-form, with conservation law $\partial _{\mu }V_{\nu
}=0$, $\tilde{\theta}$ is the Lagrange multiplier which introduce the
conservation law. The origin of this conserved quantity is the existence of
an isometry vector for the $1+D$ metric field.

From the equations of motion for the vector field $V_{\mu }$ it follows%
\begin{equation}
V_{\mu }=-\frac{1}{g_{00}}\left( g_{0a}\partial _{\mu }x^{a}+\frac{i}{\sqrt{%
\gamma }}\varepsilon _{\mu }^{~~\nu }\left( h_{0a}\partial _{\nu
}x^{a}+\partial _{\nu }\tilde{\theta}\right) \right) ,
\end{equation}%
and the Action reads%
\begin{equation}
\tilde{S}=\frac{1}{4\pi a_{0}}\int \left( \sqrt{\gamma }\gamma ^{\mu \nu
}\left( \bar{g}_{ab}\partial _{\mu }x^{\alpha }\partial _{\nu }x^{b}\right)
+i\varepsilon ^{\mu \nu }\left( \bar{h}_{ab}\partial _{\mu }x^{\alpha
}\partial _{\nu }x^{b}\right) +a_{0}\sqrt{\gamma }R^{\left( 2\right) }\bar{%
\phi}\left( x^{\gamma }\right) \right) ,  \label{dd.02a}
\end{equation}%
where%
\begin{equation}
\bar{g}_{00}=\left( g_{00}\right) ^{-1}~,~\bar{g}_{0a}=h_{0a}\left(
g_{00}\right) ^{-1}~,~\bar{g}_{ab}=g_{ab}-\left(
g_{0a}g_{0b}-h_{0a}h_{0b}\right) \left( g_{00}\right) ^{-1},
\end{equation}%
\begin{equation}
\bar{h}_{0a}=-h_{a0}=g_{0a}\left( g_{00}\right) ^{-1}~,~\bar{h}%
_{ab}=h_{ab}+\left( g_{0a}h_{0b}-h_{0a}g_{0b}\right) \left( g_{00}\right)
^{-1}.
\end{equation}%
The two Action Integrals (\ref{dd.01}), (\ref{dd.02a}) lead to the same
classical equations of motion. Moreover, the equivalency at the quantum
level leads to the conformal invariance constraint for the Action Integral (%
\ref{dd.01}), which leads to the following constraints for the dilaton field
\begin{equation}
\frac{1}{a_{0}}\frac{D-26}{48\pi ^{2}}+\frac{1}{16\pi ^{2}}\left( 4\left(
\nabla \phi \right) ^{2}-4\nabla ^{2}\phi -R+\frac{1}{12}\mathcal{H}%
^{2}\right) =0,  \label{df.01}
\end{equation}%
\begin{equation}
R_{ab}-\frac{1}{4}\mathcal{H}_{a}^{cd}\mathcal{H}_{bcd}+2\nabla _{a}\nabla
_{b}\phi =0,  \label{df.02}
\end{equation}%
\begin{equation}
\nabla _{c}\mathcal{H}_{ab}^{c}-2\left( \nabla _{c}\phi \right) \mathcal{H}%
_{ab}^{c}=0,  \label{df.03}
\end{equation}%
where $R_{ab}$ is the Ricci tensor related to the metric tensor $g_{ab}$
with Ricciscalar $R$, $\mathcal{H}_{abc}=3\nabla _{\lbrack a}\mathcal{H}%
_{bc]}$ is the antisymmetric tensor strength, and $\mathcal{H}^{2}=\mathcal{H%
}_{abc}\mathcal{H}^{abc}$ is the torsion scalar.

Moreover, the Action Integral (\ref{dd.02a}) is conformal invariant if and
only if
\begin{equation*}
\bar{\phi}=\phi -\frac{1}{2}g_{00}\text{.}
\end{equation*}

This discete transformation is characterized as duality symmetry. The origin of the
transformation is based on the existence of an isometry. In particular the origin of the transformation lies in
the $O\left( d,d\right) $ symmetry group \cite{Shapere,mm}. 

A covariant
construction of the duality transformation was presented in \cite{rvcon}.
The duality transformation relates different geometries without necessarily
having similar topological properties. For more details, we refer to the
review article \cite{rvduality}.

\section{Gasperini-Veneziano scale factor duality}

\label{sec2a}

In this Section we continue with the review of the Gasperini-Veneziano scale
factor duality symmetry and its extension in teleparallelism.

\subsection{Duality symmetry in scalar-tensor cosmology}

In \cite{gasp0,Gasp} Veneziano and Gasperini derived the duality
transformation for the dilaton field in the case of a spatially flat FLRW
geometry. Indeed, the equations of motions for the dilaton field (\ref{df.01}%
), (\ref{df.02}) and (\ref{df.03}) follows from the variation of the
following Action Integral%
\begin{equation}
S_{dilaton}=\int d^{D}x\sqrt{\left\vert g\right\vert }e^{-2\phi }\left(
R-4\nabla _{a}\phi \nabla ^{a}\phi -\frac{1}{12}\mathcal{H}^{2}-2\Lambda
\right) .  \label{d.10}
\end{equation}%
where $\Lambda $ plays the role of the cosmological constant related to the
dimension $D$ and the string tension parameter with the algebraic condition $%
\Lambda =\frac{1}{a^{\prime }}\frac{D-26}{3\pi ^{2}}$. \ \ In the following
we eliminate the antisymmetric part $\mathcal{H}^{2}=0$, and we end with the
following Action Integral
\begin{equation}
S_{ST\left( \phi \right) }=\int d^{D}x\sqrt{\left\vert g\right\vert }%
e^{-2\phi }\left( R-4\nabla _{a}\phi \nabla ^{a}\phi -2\Lambda \right) ,
\label{d.11}
\end{equation}%
which belongs to the family of the scalar-tensor theories \cite{faraonibook}%
. Without loss of generality, if we introduce the new field $\psi =-\frac{1}{%
2}\ln \phi $, the latter Action Integral (\ref{d.11}) reads%
\begin{equation}
S_{ST\left( \psi \right) }=\int d^{D}x\sqrt{\left\vert g\right\vert }\left(
\psi R-\frac{1}{\psi }\nabla _{a}\psi \nabla ^{a}\psi -2\Lambda \psi \right).
\end{equation}%
From this, we infer that $\psi$ is the Brans-Dicke field \cite{bdsc}, with a
specific value of the Brans-Dicke parameter.

We now assume that the background geometry is that of a $4$-dimensional
isotropic and homogeneous cosmology with zero spatial curvature, described
by the line element%
\begin{equation}
ds^{2}=-N(t)^{2}dt^{2}+a(t)^{2}\left( dr^{2}+r^{2}\left( d\theta ^{2}+\sin
^{2}\theta d\varphi ^{2}\right) \right) .  \label{d.13}
\end{equation}%
The field equations for the dilaton field for this geometry are%
\begin{equation}
e^{-2\phi }\left( 6a\dot{a}-12a^{2}\dot{a}\dot{\phi}+4a^{3}\dot{\phi}%
^{2}-2\Lambda \right) =0,
\end{equation}%
\begin{equation}
\ddot{a}+\frac{2}{a}\dot{a}^{2}-2\dot{a}\dot{\phi}=0,
\end{equation}%
\begin{equation}
\ddot{\phi}+\frac{3}{2a^{2}}\dot{a}^{2}-\dot{\phi}^{2}-\frac{\Lambda }{2}=0,
\end{equation}

By replacing th Ricci scalar $R=6\left( \frac{\ddot{a}}{a}+\left( \frac{\dot{%
a}}{a}\right) ^{2}\right) $ in (\ref{d.11}) and integrating by parts we
derive the point-like Lagrangian%
\begin{equation}
L\left( a,\dot{a},\phi ,\dot{\phi}\right) =e^{-2\phi }\left( 6a\dot{a}%
^{2}-12a^{2}\dot{a}\dot{\phi}+4a^{3}\dot{\phi}^{2}-2\Lambda a^{3}\right) .
\end{equation}%
Under the change of variables
\begin{equation}
b=a^{-1}~,~\zeta =\phi -3\ln a,  \label{d.13b}
\end{equation}%
we end with the point-like Lagrangian
\begin{equation}
\bar{L}\left( b,\dot{b},\zeta ,\dot{\zeta}\right) =e^{-2\zeta }\left( 6b\dot{%
b}^{2}-12b^{2}b\dot{\zeta}+4b^{3}\dot{\zeta}^{2}-\Lambda b^{3}\right) ,
\end{equation}%
which leads to the same classical field equations.

Transformation (\ref{d.13b}) is the Gasperini-Veneziano duality symmetry,
and it can connect the solutions for the early and the late-time universes.

Although the duality symmetry occurs on the background level, it can be used
to study cosmological perturbations \cite{gasp1}. The presence of the
duality symmetry, specifically that of the $O\left( d,d\right) $ symmetry,
is essential for the existence of solutions to the Wheeler-DeWitt equation
of quantum cosmology \cite{keha}. The continuous symmetry that generates the
duality transformation was calculated using the Noether symmetry in \cite%
{ans}. Moreover, in \cite{ang}, the nature of the duality symmetry was
investigated in conformal equivalent theories to the Action Integral (\ref%
{d.11}), while in \cite{dualc}, cyclic universes with duality scale factor
symmetry were investigated.

\subsection{Duality symmetry in teleparallelism}

In teleparallelism \cite{tl0,tl1,tl3,tl4} the fundamental scalar is the
torsion $T$ constructed by the vierbein fields.

In \cite{pOdd} it was found that in a four-dimensional manifold the
teleparallel dark energy model with Action Integral
\begin{equation}
S_{T}=\frac{1}{16\pi G}\int d^{4}xe\left[ e^{-2\phi }\left( T+\frac{\omega }{%
2}\phi _{;\mu }\phi ^{\mu }-2\Lambda \right) \right] ,  \label{d.14}
\end{equation}%
within a spatially flat FLRW geometry (\ref{d.13}) possesses a discrete
symmetry which is of the same origin with that of the Gasperini-Veneziano
scale factor duality symmetry.

For the line element (\ref{d.13}) and for diagonal vierbein fields the
torsion scalar $T$ is derived $T=6\left( \frac{\dot{a}}{a}\right) ^{2}$, and
the point-like Lagrangian which follows from (\ref{d.14}) is defined as
\begin{equation}
L\left( a,\dot{a},\phi ,\dot{\phi}\right) =e^{-2\phi }\left( 6a\dot{a}^{2}-%
\frac{\omega }{2}a^{3}\dot{\phi}^{2}-2\Lambda a^{3}\right) .  \label{d.15}
\end{equation}%
The gravitational field equations are%
\begin{equation}
\left( 6\left( \frac{\dot{a}}{a}\right) ^{2}-\frac{\omega }{2}\dot{\phi}%
^{2}+2\Lambda \right) =0~,
\end{equation}%
\begin{equation}
\frac{\ddot{a}}{a}+2\left( \frac{\dot{a}}{a}\right) ^{2}+\frac{1}{2}\left(
\frac{\omega }{2}\dot{\phi}^{2}+2\Lambda \right) -4\left( \frac{\dot{a}}{a}%
\right) \dot{\phi}=0~,  \label{d.12}
\end{equation}%
\begin{equation}
\omega \left( \ddot{\phi}+3\left( \frac{\dot{a}}{a}\right) \dot{\phi}\right)
+\left( \frac{\omega }{2}\dot{\phi}^{2}-2\Lambda \right) =0~.
\end{equation}

Thus, the transformation
\begin{equation}
a\rightarrow a^{p_{1}}e^{p_{2}\phi }~,~\phi \rightarrow p_{4}\phi +p_{3}\ln
a,  \label{d.16a}
\end{equation}%
with%
\begin{equation}
p_{1}=\frac{1+\kappa ^{2}}{1-\kappa ^{2}}~,~p_{2}=-\frac{4}{3\left( 1-\kappa
^{2}\right) }~,~p_{3}=\frac{3\kappa ^{2}}{1-\kappa ^{2}}~,~p_{4}=-\frac{%
1+\kappa ^{2}}{1-\kappa ^{2}}~,~\omega =\frac{16}{3\kappa ^{2}}.
\label{d.16}
\end{equation}%
leaves invariant the point-like Lagrangian function (\ref{d.15}),
consequently, the resulting field equations. The explicitly form of the
Gasperini-Veneziano scale-factor duality transformation is recovered for
large values of parameter $\kappa $, that is, $\omega \rightarrow 0$. Indeed
when $\kappa \rightarrow \infty $ it follows%
\begin{equation}
p_{1}=-1~,~p_{2}=0~,~p_{3}=3~,~p_{4}=1~,~\omega =0.
\end{equation}

On the other hand, for $\kappa ^{2}\rightarrow 1$, the discrete
transformation is not defined. Similarly with the case of the scalar-tensor
theory, the origin of the transformation in teleparallel dark energy is the
existence of the $O\left( d,d\right) $ symmetry \cite{pOdd} for the
gravitational model.

The study of the asymptotics for this teleparallel dark energy model \cite%
{anst1} gives that this model unifies the early and late time acceleration
eras of the universe.

\section{Symmetric teleparallel FLRW Cosmology}

\label{sec3}

In the framework of symmetric teleparallel theory we introduce a scalar
field nonminimally coupled such that the gravitational Action Integral to be
\cite{sc1,gg1}
\begin{equation}
S_{ST\varphi }=\int d^{4}x\sqrt{-g}\left( \varphi Q-\frac{\omega \left(
\varphi \right) }{2}g^{\mu \nu }\varphi _{,\mu }\varphi _{,\nu }-V\left(
\varphi \right) \right) ,  \label{st.01}
\end{equation}%
where $\varphi $ is the scalar field, $V\left( \varphi \right) $ is the
scalar field potential which defines the mass and function $\omega \left(
\varphi \right) $ defines the coupling between the scalar field and the
geometric scalar $Q$. \ The latter Action Integral defines the symmetric
teleparallel scalar-tensor theory which properties similar wit a Machian
theory \cite{revmach}, either if it is not pure Machian theory \cite{sc1},
it has properties similar to that of a Machian theory.

Scalar $Q$, is the nonmetricity scalar constructed by the symmetric and flat
connection $\Gamma _{\mu \nu }^{\kappa }$ used to define covariant
derivative $\nabla _{\mu }$ in the four-dimensional manifold with metric $%
g_{\mu \nu }$. It holds,
\begin{equation}
\nabla _{\kappa }g_{\mu \nu }=Q_{\kappa \mu \nu },
\end{equation}%
and%
\begin{equation}
Q=Q_{\lambda \mu \nu }P^{\lambda \mu \nu }
\end{equation}%
in which $P^{\lambda \mu \nu }$ is defined as%
\begin{equation}
P_{~\mu \nu }^{\lambda }=\frac{1}{4}\left( -2L_{~~\mu \nu }^{\lambda
}+Q^{\lambda }g_{\mu \nu }-Q^{\prime \lambda }g_{\mu \nu }-\delta _{(\mu
}^{\lambda }Q_{\nu )}\right)
\end{equation}%
where $L_{~\mu \nu }^{\lambda }=\frac{1}{2}g^{\lambda \sigma }\left( Q_{\mu
\nu \sigma }+Q_{\nu \mu \sigma }-Q_{\sigma \mu \nu }\right) ~$and $%
Q_{\lambda }=Q_{\lambda ~~~\mu }^{~~~\mu },Q_{\lambda }^{\prime
}=Q_{~~\lambda \mu }^{\mu }$.

The gravitational field equation follows from the variation of (\ref{st.01})
with respect to the metric tensor, the scalar field $\varphi $ and the
connection. Indeed, they are

\begin{equation}
\varphi G_{\mu \nu }+2\varphi _{,\lambda }P_{~~\mu \nu }^{\lambda }+g_{\mu
\nu }V\left( \varphi \right) +\frac{\omega \left( \varphi \right) }{2}\left(
g_{\mu \nu }g^{\lambda \kappa }\varphi _{,\lambda }\varphi _{,\kappa
}-\varphi _{,\mu }\varphi _{,\nu }\right) =0.  \label{rs.18}
\end{equation}%
\begin{equation}
\frac{\omega \left( \varphi \right) }{\sqrt{-g}}g^{\mu \nu }\partial _{\mu
}\left( \sqrt{-g}\partial _{\nu }\varphi \right) +\frac{\omega _{,\varphi }}{%
2}g^{\lambda \kappa }\varphi _{,\lambda }\varphi _{,\kappa }+\varphi
Q-V_{,\varphi }=0,  \label{rs.19}
\end{equation}
\begin{equation}
\nabla _{\mu }\nabla _{\nu }\left( \sqrt{-g}\varphi P_{~~~~\sigma }^{\mu \nu
}\right) =0.  \label{rs.20}
\end{equation}

At this point we remark that for $\omega \left( \varphi \right) =0$, the
Action Integral (\ref{st.01}) is equivalent with the Action Integral of $%
f\left( Q\right) $-gravity, that is~\cite{fq1,fq2,fq3,fq4,fq5}
\begin{equation}
S_{f\left( Q\right) }=\int d^{4}x\sqrt{-g}f\left( Q\right) ,
\end{equation}%
where the scalar field $\varphi $ and the potential function $V\left(
\varphi \right) $ are related with the function $f\left( Q\right) $, via the
relations $\varphi =f^{\prime }\left( Q\right) $ and $V\left( \varphi
\right) =f_{,Q}Q-f~~$\cite{palf1}.

On the other hand, for $\omega \left( \varphi \right) =\frac{\omega _{0}}{%
4\varphi }$, the model (\ref{st.01}) has been characterized as the
Brans-Dicke analogue in symmetric teleparallel theory \cite{palf3}. Indeed,
for this specific coupling function we can define the new scalar $\varphi
=e^{-2\phi }$, such that, Lagrangian (\ref{st.01}) to expressed as follows%
\begin{equation}
S_{D\left( Q\right) }=\int d^{4}x\sqrt{-g}e^{-2\phi }\left( Q-\frac{\omega
_{0}}{2}g^{\mu \nu }\phi _{,\mu }\phi _{,\nu }-\hat{V}\left( \phi \right)
\right) ~,~\hat{V}\left( \phi \right) =V\left( \phi \right) e^{-2\phi }\ ,
\label{rs.21}
\end{equation}%
where now $\phi $ describes the dilaton field in symmetric teleparallel
theory.

\subsection{FLRW Cosmology}

We assume an isotropic and homogeneous universe described by the spatially
flat FLRW geometry. For this spacetime, the requirements the connection $%
\Gamma _{\mu \nu }^{\kappa }$ to be symmetric, flat and to inherits the
symmetries of the spacetime leads to three families of connections \cite%
{Heis2,ndcon}, the $\Gamma ^{A}$,~$\Gamma ^{B}$ and $\Gamma ^{C}$.

For the metric tensor we consider the coordinate system $\left( t,r,\theta
,\phi \right) $, where the line element is expressed as in the relation (\ref%
{d.13}), then the common nonzero coefficients for the three families of
connections are \cite{Heis2,ndcon}
\begin{align*}
\Gamma _{\theta \theta }^{r}& =-rr~,~\Gamma _{~\varphi \varphi }^{r}=-r\sin
^{2}\theta , \\
\Gamma _{\;r\theta }^{\theta }& =\Gamma _{\;\theta r}^{\theta }=\Gamma
_{\;r\varphi }^{\varphi }=\Gamma _{\;\varphi r}^{\varphi }=\frac{1}{r}~,~ \\
\Gamma _{\varphi \varphi }^{\theta }& =-\sin \theta \cos \theta ~,~\Gamma
_{\theta \varphi }^{\varphi }=\Gamma _{\varphi \theta }^{\varphi }=\cot
\theta .
\end{align*}%
while for each connection we have the additional nonzero components%
\begin{equation*}
\Gamma ^{A}:\Gamma _{\;tt}^{t}=\gamma (t)
\end{equation*}%
\begin{equation*}
\Gamma ^{B}:\Gamma _{\;tt}^{t}=\frac{\ddot{\psi}(t)}{\dot{\psi}(t)}+\dot{\psi%
}(t),\quad \Gamma _{\;tr}^{r}=\Gamma _{\;rt}^{r}=\Gamma _{\;t\theta
}^{\theta }=\Gamma _{\;\theta t}^{\theta }=\Gamma _{\;t\varphi }^{\varphi
}=\Gamma _{\;\varphi t}^{\varphi }=\gamma (t),
\end{equation*}%
and%
\begin{equation*}
\Gamma ^{C}:\Gamma _{\;tt}^{t}=\frac{\ddot{\Psi}}{\dot{\Psi}},\quad \Gamma
_{\;rr}^{t}=\frac{1}{\dot{\Psi}},\quad \Gamma _{\;\theta \theta }^{t}=\frac{%
r^{2}}{\dot{\Psi}},\quad \Gamma _{\;\varphi \varphi }^{t}=\frac{r^{2}\sin
^{2}\theta }{\dot{\Psi}}.
\end{equation*}

The scalar $\gamma \left( t\right) $,~$\psi \left( t\right) $ and $\Psi
\left( t\right) $, introduce geometric degrees of freedom in the
gravitational model and they are constraint by the equation of motion (\ref%
{rs.20}). We remark that that in symmetric teleparallel gravity the
selection of the connection is not unique The physical viability of the
theory is still under debate \cite{cc0,cc1,cc2,cc3}.

For the connection $\Gamma ^{A}$ the equation of motion (\ref{rs.20}) is
trivially satisfied, and $\gamma \left( t\right) $ plays no role in the
gravitational dynamics, that is, connection $\Gamma ^{A}$ is defined in the
so-called coincidence gauge. Although the other two connections $\Gamma ^{B}$
and $\Gamma ^{C}$ are flat, there exist a coordinate system where they have
zero components, the selection of the line element (\ref{d.13}) constraints
the connections.

For each of the above connections we calculate a different nonmetricity
scalar $Q$. Specifically, the three scalars are%
\begin{eqnarray}
Q\left( \Gamma ^{A}\right) &=&-6H^{2}, \\
Q\left( \Gamma ^{B}\right) &=&-6H^{2}+\frac{3}{a^{3}}\left( a^{3}\psi
\right) ^{\cdot }~, \\
Q\left( \Gamma ^{C}\right) &=&-6H^{2}+\frac{3}{a^{3}}\left( \frac{a}{\Psi }%
\right) ^{\cdot }~.
\end{eqnarray}

Therefore, by replacing in (\ref{rs.21}) we end up with three point-like
Lagrangians, and three different sets of dynamical systems.

From the connection $\Gamma ^{A}$ the point-like Lagrangian reads%
\begin{equation}
L^{A}\left( a,\dot{a},\phi ,\dot{\phi}\right) =e^{-2\phi }\left( 6\phi a\dot{%
a}^{2}-\frac{\omega _{0}}{2}a^{3}\dot{\phi}^{2}-a^{3}\hat{V}\left( \phi
\right) \right) ,  \label{ln.01}
\end{equation}%
for connection $\Gamma ^{B}~$we calculate the point-like Lagrangian function%
\begin{equation}
L^{B}\left( a,\dot{a},\phi ,\dot{\phi},\psi ,\dot{\psi}\right) =e^{-2\phi
}\left( 6a\dot{a}^{2}-\frac{\omega _{0}}{2}a^{3}\dot{\phi}^{2}+3a^{3}\dot{%
\phi}\dot{\psi}-a^{3}\hat{V}\left( \phi \right) \right) ,  \label{ln.02}
\end{equation}%
while for connection $\Gamma ^{C}$ we derive the point-like Lagrangian\qquad
\begin{equation}
L^{C}\left( a,\dot{a},\phi ,\dot{\phi},\psi ,\dot{\psi}\right) =e^{-2\phi
}\left( 6a\dot{a}^{2}-\frac{\omega _{0}}{2}a^{3}\dot{\phi}^{2}+\frac{3}{a^{2}%
}\frac{\dot{\phi}}{\dot{\Psi}}-a^{3}\hat{V}\left( \phi \right) \right) .
\label{ln.03}
\end{equation}%
The gravitational field equations for the latter cosmological models can be
found in \cite{ansf11}, with a small change on the parameter $\omega
_{0}\rightarrow -\omega _{0}$ and $\phi \rightarrow -\frac{1}{2}\phi $.

\section{Generalized duality symmetry}

\label{sec4}

In this Section we investigate the existence of scale factor duality
symmetries for the dynamical systems described by the three point-like
Lagrangian functions (\ref{ln.01}), (\ref{ln.02}) and (\ref{ln.03}). For the
potential function we assume $\hat{V}\left( \phi \right) =2\Lambda $.

We observe that Lagrangian $L^{A}$ (\ref{ln.01}) is equivalent with
Lagrangian function (\ref{d.15}) of teleparallel gravity, consequently the
same generalized scale factor duality transformation given by expressions (%
\ref{d.16a}) and (\ref{d.16}) is valid.

Furthermore, for the Lagrangian $L^{B}~$(\ref{ln.02}) we find derive the
generalized scale factor duality symmetry with discrete transformation
\begin{equation}
a\rightarrow a^{p_{1}}e^{p_{2}\phi }e^{\frac{3}{8}p_{7}\psi }~,~\phi
\rightarrow p_{4}\phi +p_{3}\ln a+\frac{9}{16}p_{7}\psi ,~\psi \rightarrow
p_{5}\ln a-p_{6}\psi
\end{equation}%
with%
\begin{eqnarray*}
p_{1} &=&\frac{1-\kappa ^{2}}{1+\kappa ^{2}},~p_{2}=0,~p_{3}=-3+\frac{3}{%
1+\kappa ^{2}},~p_{4}=1,~p_{5}=-\frac{16}{3\left( 1+\kappa ^{2}\right) },~ \\
p_{6} &=&-1+\frac{2}{1+\kappa ^{2}},~p_{7}=-\frac{2\kappa ^{2}}{1+\kappa ^{2}%
},~\omega _{0}=\frac{16}{3\kappa ^{2}}
\end{eqnarray*}%
or%
\begin{eqnarray*}
p_{1} &=&\frac{1+\kappa ^{2}}{1-\kappa ^{2}},~p_{2}=-\frac{4}{3\left(
1-\kappa ^{2}\right) },~p_{3}=\frac{3\kappa ^{2}}{1-\kappa ^{2}},~p_{4}=-%
\frac{1+\kappa ^{2}}{1-\kappa ^{2}},~ \\
p_{5} &=&0,~p_{6}=1,~p_{7}=\frac{2\kappa ^{2}}{1-\kappa ^{2}},~\omega _{0}=%
\frac{16}{3\kappa ^{2}}.
\end{eqnarray*}

We observe that there are two discrete transformations which leave
Lagrangian $L^{B}$ invariant. In the extreme limit where $\omega _{0}=0$,
that is,~$\kappa \rightarrow +\infty $, the two transformations have the
limits%
\begin{equation}
a\rightarrow a^{-1}e^{-\frac{3}{4}\psi }~,~\phi \rightarrow \phi -3\ln a-%
\frac{9}{8}\psi ,~\psi \rightarrow -\frac{1}{2}\psi ,
\end{equation}%
and%
\begin{equation}
a\rightarrow a^{-1}e^{-\frac{3}{4}\psi }~,~\phi \rightarrow \phi -3\ln a+%
\frac{9}{8}\psi ,~\psi \rightarrow \psi .
\end{equation}

Finally, for the third Lagrangian function $L^{C}$ (\ref{ln.03}) we find the
generalized scale factor duality symmetry with discrete transformation%
\begin{equation}
a\rightarrow a^{p_{1}}e^{p_{2}\phi }f,~\phi \rightarrow p_{4}\phi +p_{3}\ln
a+\frac{3}{2}\ln f,~\dot{\Psi}\rightarrow a^{p_{5}}e^{p_{6}\phi }B\left(
\dot{\Psi}\right) ,
\end{equation}%
where%
\begin{eqnarray*}
p_{1} &=&\frac{1+\kappa ^{2}}{1-\kappa ^{2}},~p_{2}=-\frac{4}{3\left(
1-\kappa ^{2}\right) },~p_{3}=\frac{3\kappa ^{2}}{\left( 1-\kappa
^{2}\right) },~ \\
p_{4} &=&-\frac{1+\kappa ^{2}}{1-\kappa ^{2}},~~p_{5}=-5\frac{1+\kappa ^{2}}{%
1-\kappa ^{2}},~p_{6}=\frac{20}{3\left( 1-\kappa ^{2}\right) },~\omega _{0}=%
\frac{16}{3\kappa ^{2}}
\end{eqnarray*}%
and functions~\thinspace $f\left( t\right) $ is defined as
\begin{equation}
f\left( t\right) =-5^{\frac{1}{5}}\frac{3\kappa ^{2}}{4\left( 1-\kappa
^{2}\right) }\int \frac{1}{B\left( \dot{\Psi}\right) }dt,
\end{equation}%
and $5^{\frac{3}{5}}B\left( t\right) f\left( t\right) =\dot{\Psi}$.

Similarly as before in the limit where $\kappa \rightarrow +\infty $, that
is, $\omega _{0}=0$, the latter discrete transformation reads%
\begin{equation}
a\rightarrow a^{-1}\exp \left( -\frac{3}{4}\int \frac{1}{\dot{\Psi}}%
dt\right) ,~\phi \rightarrow \phi -\ln a-\frac{9}{8}\int \frac{1}{\dot{\Psi}}%
dt,~\Psi \rightarrow a^{5}\Psi .
\end{equation}

The generalized scale factor duality symmetry for the Lagrangian function $%
L^{C}$ it is a nonlocal discrete transformation. It is different from the
discrete transformations for the other two Lagrangians, namely $L^{A}$ and $%
L^{B}$ where the origin of the transformation are point symmetries which
correspond to to local transformations.

\section{Integrability for the cosmological model}

\label{sec5}

For the scalar-tensor (\ref{d.11}) and the scalar-torsion (\ref{d.14})
theories, the origin of the discrete transformations are local continuous
transformations \cite{ang,pOdd}. From Noether's theorem \cite{revs} it is
known that there exist a conserved quantity for any continuous
transformation which keeps the variation of the Action Integral invariant.
Thus, in the following lines we investigate the continuous transformations
which keep invariant the cosmological Lagrangians for the symmetric
teleparallel model of our consideration.

The connection defined in the coincidence gauge leads to a cosmological
model\ with Lagrangian (\ref{ln.01}) where the background field equations
are equivalent to the scalar-torsion theory (\ref{d.14}) the analysis
presented in \cite{pOdd} is valid. Therefore, we focus to the analysis of
Lagrangians (\ref{ln.02}) and (\ref{ln.03}).

\subsection{Conservation laws for connection $\Gamma ^{B}$}

For Lagrangian function (\ref{ln.02}), namely $L^{B}$ we determine the
conservation laws%
\begin{eqnarray}
I_{1}^{B} &=&e^{-2\phi }\left( 6a\dot{a}^{2}-\frac{\omega _{0}}{2}a^{3}\dot{%
\phi}^{2}+3a^{3}\dot{\phi}\dot{\psi}+2\Lambda a^{3}\right) , \\
I_{2}^{B} &=&e^{-2\phi }a\left( 24\dot{a}-3a\left( \omega _{0}+\frac{16}{3}%
\right) \dot{\phi}+9a\dot{\psi}\right) , \\
I_{3}^{B} &=&e^{-2\phi }a^{3}\dot{\phi}, \\
I_{4}^{B} &=&e^{-2\phi }a^{2}\left( 8\phi \dot{\alpha}+3a\phi \dot{\psi}%
-a\left( 3\psi +8\ln a\right) \dot{\phi}\right) , \\
I_{5}^{B} &=&\frac{\sqrt{3}\cos \left( \sqrt{3\Lambda }t\right) }{2\sqrt{%
\Lambda }}I_{1}^{B}+4e^{-2\phi }\sin \left( \sqrt{3\Lambda }t\right) \left(
3a^{2}\dot{a}-2a^{3}\dot{\phi}\right) , \\
I_{6}^{B} &=&-\frac{\sqrt{3}\sin \left( \sqrt{3\Lambda }t\right) }{\sqrt{%
\Lambda }}I_{1}^{B}+4e^{-2\phi }\cos \left( \sqrt{3\Lambda }t\right) \left(
3a^{2}\dot{a}-2a^{3}\dot{\phi}\right) .
\end{eqnarray}

Conservation law $I_{1}^{B}$ is the Hamiltonian for the autonomous system (%
\ref{ln.02}) and it follows from the symmetry vector $X_{1}^{B}=\partial
_{t} $ for the Lagrangian function. Due to the constraint equation \cite%
{ansf11}, value $I_{I}^{B}=0$.

The conserved quantity $I_{2}^{B}$ is generated by the transformation with
infinitesimal transformation.
\begin{equation*}
X_{1}^{B}=2a\partial _{a}+3\partial _{\phi }+\left( \frac{4}{\kappa ^{2}}-%
\frac{16}{3}\right) \partial _{\psi },~\omega _{0}=\frac{16}{3\kappa ^{2}}
\end{equation*}%
This is the generator of the discrete generalized scale-factor duality
symmetry derived before.

Conservation law $I_{3}^{B}$ follows from the vector field~%
\begin{equation}
X_{1}^{B}=\partial _{\psi },
\end{equation}
while $I_{4}^{B}$ from the vector field
\begin{equation}
X_{4}^{B}=\frac{2}{3}a\phi \partial _{a}+\phi \partial _{\phi }+\frac{8}{3}%
\left( \frac{2}{3\kappa ^{2}}\phi -\ln a\right) \partial _{\psi },~\omega
_{0}=\frac{16}{3\kappa ^{2}}
\end{equation}

Finally, the two conservation laws~$I_{5}^{B}$, and $I_{6}^{B}$ are due to
the generators\qquad\
\begin{equation}
X_{5}^{B}=\frac{\sqrt{3}\cos \left( \sqrt{3\Lambda }t\right) }{\sqrt{\Lambda
}}\partial _{t}-\sin \left( \sqrt{3\Lambda }t\right) \left( a\partial _{a}-%
\frac{8}{3}\partial _{\psi }\right) ,
\end{equation}
and
\begin{equation}
X_{6}^{B}=\frac{\sqrt{3}\sin \left( \sqrt{3\Lambda }t\right) }{\sqrt{\Lambda
}}\partial _{t}+\cos \left( \sqrt{3\Lambda }t\right) \left( a\partial _{a}-%
\frac{8}{3}\partial _{\psi }\right) .
\end{equation}%
However, because of the constraint equation these two conservation laws
reads~$I_{5}^{B}=4e^{-2\phi }\sin \left( \sqrt{3\Lambda }t\right) \left(
3a^{2}\dot{a}-2a^{3}\dot{\phi}\right) $ and $I_{6}^{B}=4e^{-2\phi }\cos
\left( \sqrt{3\Lambda }t\right) \left( 3a^{2}\dot{a}-2a^{3}\dot{\phi}\right)
.$

We conclude that the dynamical system described by the Lagrangian function (%
\ref{ln.02}) is superintegrable.

\subsection{Conservation laws for connection $\Gamma ^{C}$}

For the Lagrangian function (\ref{ln.03}), namely $L^{C}$, which corresponds
to the third connection we calculate the conserved quantities%
\begin{eqnarray}
I_{1}^{C} &=&e^{-2\phi }\left( 6a\dot{a}^{2}-\frac{\omega _{0}}{2}a^{3}\dot{%
\phi}^{2}-\frac{3}{a^{2}}\frac{\dot{\phi}}{\dot{\Psi}}+2\Lambda a^{3}\right)
, \\
I_{2}^{C} &=&e^{-2\phi }\left( 8a^{2}\dot{a}-\omega _{0}a^{3}\dot{\phi}+%
\frac{3}{a^{2}}\frac{1}{\dot{\Psi}}+\frac{10\Psi }{a^{2}}\frac{\dot{\phi}}{%
\dot{\Psi}^{2}}\right) , \\
I_{3}^{C} &=&\frac{e^{-2\phi }}{a^{2}}\frac{\dot{\phi}}{\dot{\Psi}^{2}}.
\end{eqnarray}%
From the field equation \cite{ansf11}, it follows that the Hamiltonian is
zero, i.e. $I_{1}^{C}=0$. These three conservation laws are generated by the
vector fields $X_{1}^{C}=\partial _{t}$,~\ \
\begin{equation}
X_{2}^{C}=\frac{2}{3}\partial _{a}+\partial _{\phi }+\frac{10}{3}\Psi
\partial _{\Psi }\text{,}
\end{equation}%
and%
\begin{equation}
X_{3}^{C}=\partial _{\Psi }\text{.}
\end{equation}

The vector field $X_{2}^{C}$ it is the generator of the discrete
transformation. Although the three conservation laws $I_{1}^{C}\,,~I_{2}^{C}%
\,$\ and $I_{3}^{C}$ are independent, they are not in involution; hence, we
cannot infer about the integrability.

\section{Analytic solution for connection $\Gamma ^{B}$}

\label{sec5a}

We employ the Hamilton-Jacobi method to solve the field equations for the
superintegrable cosmological model described by the Lagrangian function (\ref%
{ln.02}).

We define the momentum%
\begin{eqnarray*}
p_{a} &=&12e^{-2\phi }a\dot{a},~ \\
p_{\phi } &=&a^{3}e^{-2\phi }\left( -\frac{16}{3\kappa ^{2}}\dot{\phi}+3\dot{%
\psi}\right) ,~ \\
p_{\psi } &=&3a^{3}e^{-2\phi }\dot{\phi}.
\end{eqnarray*}%
Therefore,%
\begin{equation}
\dot{a}=\frac{e^{2\phi }}{12a}p_{a}~,~\dot{\phi}=\frac{e^{2\phi }}{3a^{3}}%
p_{\psi },~\dot{\psi}=\frac{e^{2\phi }}{27\kappa ^{2}a^{3}}\left( 16p_{\psi
}+9p_{\phi }\kappa ^{2}\right) .
\end{equation}

From the conservation laws $I_{1}^{B}$,~$I_{2}^{B}$ and $I_{3}^{B}$ we find
the solution for the Hamilton-Jacobi equation; that is, the function form
for the action $S\left( a,\phi ,\psi \right) $ reads%
\begin{eqnarray}
S\left( a,\phi ,\psi \right) &=&+I_{3}^{B}\left( \psi +\frac{8}{3}\ln
a\right) +\frac{3I_{2}^{B}\kappa ^{2}-12I_{3}^{B}}{9\kappa ^{2}}\phi  \notag
\\
&&-\frac{4i}{9\kappa ^{2}}\int \left( \sqrt{108e^{-4\phi }\Lambda
a^{6}\kappa ^{4}+6\kappa ^{2}I_{3}^{B}\left( \left( \frac{8}{3}%
I_{3}^{B}+I_{2}^{B}\right) \kappa ^{2}+\frac{4}{3}I_{3}^{B}\right) }\right)
d\phi
\end{eqnarray}

Hence, the reduced classical field equations are
\begin{eqnarray}
\dot{a} &=&\frac{e^{2\phi }}{18\kappa ^{2}a^{2}}\sqrt{4I_{3}^{B}\kappa
^{2}\left( 2I_{3}^{B}-\left( 3I_{2}^{B}+8I_{3}^{B}\right) \kappa ^{2}\right)
-108a^{6}e^{-4\phi }\kappa ^{4}\Lambda },  \label{dd1} \\
\dot{\phi} &=&\frac{e^{2\phi }}{3a^{3}}I_{3}^{B}  \label{dd2} \\
\dot{\psi} &=&\frac{e^{2\phi }}{27\kappa ^{2}a^{3}}\left( 3I_{2}^{B}\kappa
^{2}+4I_{3}^{B}\left( 1+4\kappa ^{2}\right) -4\sqrt{4I_{3}^{B}\kappa
^{2}\left( 2I_{3}^{B}-\left( 3I_{2}^{B}+8I_{3}^{B}\right) \kappa ^{2}\right)
-108a^{6}e^{-4\phi }\kappa ^{4}\Lambda }\right) .
\end{eqnarray}%
We remark that the latter dynamical system is decomposable and the evolution
of the scalar $\psi $ plays no role in the physical space.

For zero value of the cosmological constant, i.e. $\Lambda =0$,\ or when
term $4I_{3}^{B}\kappa ^{2}\left( 2I_{3}^{B}-\left(
3I_{2}^{B}+8I_{3}^{B}\right) \kappa ^{2}\right) $ dominates it follows%
\begin{equation*}
\frac{\sqrt{4I_{3}^{B}\kappa ^{2}\left( 2I_{3}^{B}-\left(
3I_{2}^{B}+8I_{3}^{B}\right) \kappa ^{2}\right) }}{18\kappa ^{2}}\int d\phi =%
\frac{I_{3}^{B}}{3}\frac{da}{a},
\end{equation*}%
that is%
\begin{equation}
\phi \left( a\right) =\frac{I_{3}^{B}}{3}\frac{18\kappa ^{2}}{\sqrt{%
4I_{3}^{B}\kappa ^{2}\left( 2I_{3}^{B}-\left( 3I_{2}^{B}+8I_{3}^{B}\right)
\kappa ^{2}\right) }}\ln a=C_{1}\left( \kappa ,I_{2}^{B},I_{1}^{B}\right)
\ln a
\end{equation}%
and%
\begin{equation}
\psi \left( a\right) =C_{2}\left( \kappa ,I_{2}^{B},I_{1}^{B}\right) \ln a.
\end{equation}

Consequently, the Hubble function reads $H\left( a\right) \simeq
a^{2C_{1}\left( \kappa ,I_{2}^{B},I_{1}^{B}\right) -3}$, which describes a
universe dominated by an ideal gas.

On the other hand when $4I_{3}^{B}\kappa ^{2}\left( 2I_{3}^{B}-\left(
3I_{2}^{B}+8I_{3}^{B}\right) \kappa ^{2}\right) =0$, that is, the $\Lambda $
term dominates we calculate%
\begin{equation}
\phi \left( a\right) =-\frac{1}{2}\ln \left( 2\left( \frac{I_{3}^{B}}{18%
\sqrt{3}a^{3}\kappa ^{2}\sqrt{-\Lambda }}-\phi _{1}\right) \right) ,
\end{equation}%
that is, the Hubble function is found to be a constant $H=const$.

We calculated the two asymptotic behaviours for the universe using this
solution. We see that a scaling solution and the de Sitter universe are
provided by the model. This result is in agreement with the general
behaviour of the theory, as analyzed in \cite{palf1,palf3}.

\begin{figure}[tbph]
\centering\includegraphics[width=1\textwidth]{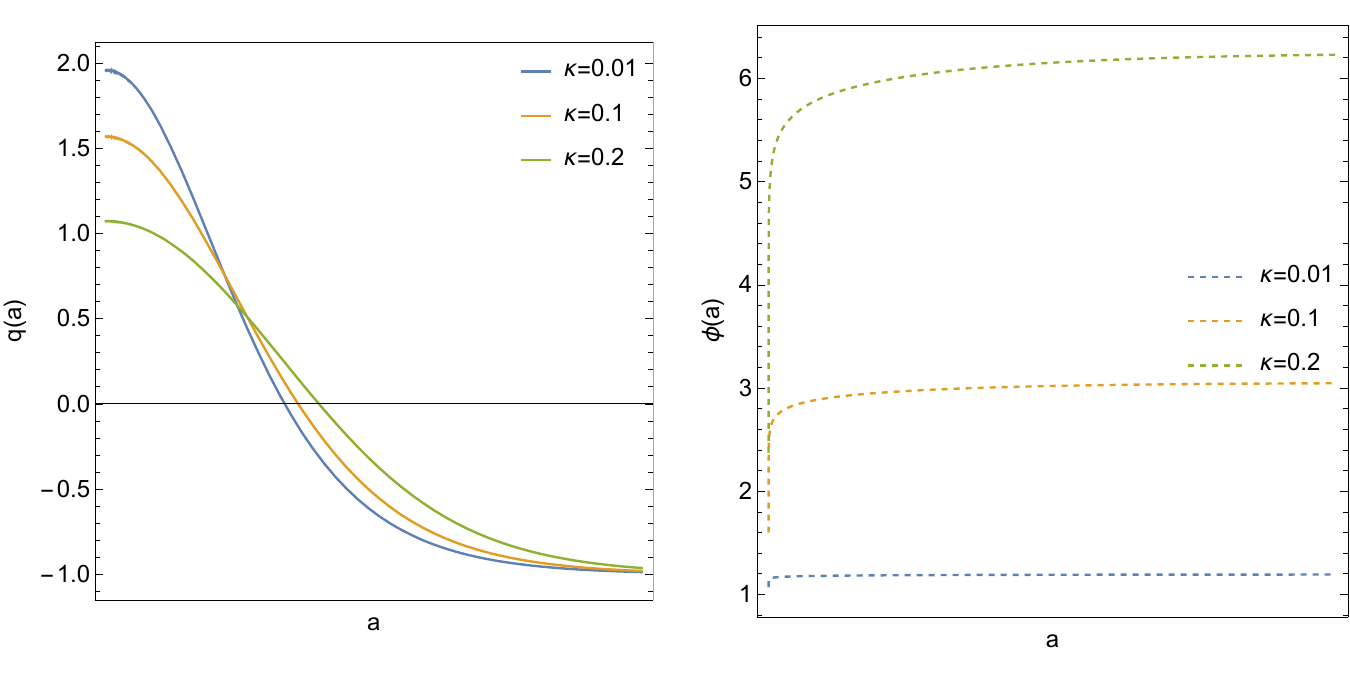}
\caption{Qualitative evolution for the deceleration parameter $q\left(
a\right) ~$(Left Fig.) and for the scalar field $\protect\phi \left(
a\right) ~$(Right Fig.) as they are given by the numerical solution for the
two-dimensional dynamical system (\protect\ref{dd1}), (\protect\ref{dd2}).
Figures are for negative value of the cosmological constant $\Lambda $ and
for positive value for the parameter $\protect\omega _{0}$. The de Sitter
universe is an attractor for the dynamical system, while the early universe
is described by the scaling solution.}
\label{fig1}
\end{figure}

In Fig. \ref{fig1}, we present the qualitative evolution of the scalar field
$\phi$ and the deceleration parameter $q=-\frac{\ddot{a}a}{\dot{a}^{2}}$, as
determined by the numerical solution of the two-dimensional dynamical system
(\ref{dd1}), (\ref{dd2}). The plots are for a negative value of the
cosmological constant $\Lambda$ and a positive value of the parameter $%
\omega_{0}$. We remark that the de Sitter universe is an attractor for the
dynamical system, while the early universe is described by a scaling
solution.

\section{\protect\bigskip}

\section{Conclusions}

\label{sec6}

Discrete transformations, which are generalized scale-factor duality
symmetries, are studied for the dilaton gravitational Action Integral in the
trinity of gravity. The Gasperini-Veneziano scale-factor duality
transformation for the scalar-tensor theory is examined within the
teleparallelism framework, in the scalar-torsion model, and in the symmetric
teleparallel scalar-tensor model. In these theories, the dilaton field is
nonminimally coupled to the Ricci scalar $R$, the torsion scalar $T$, or the
nonmetricity scalar $Q$, respectively.

In teleparallel and symmetric teleparallel theories, we found families of
generalized Gasperini-Veneziano scale-factor duality transformations which
keep the gravitational Lagrangians invariant. The Gasperini-Veneziano
transformation exists as a limit when the coupling term of the gravitational
Lagrangian dominates. Furthermore, we examined the existence of continuous
transformations that generate the discrete transformations, and by using
Noether's theorem, we determined the corresponding conservation laws for the
dilaton Action in the trinity of gravity.

Finally, the integrability properties of the cosmological models
investigated, and we derived a new analytic solution within the symmetric
teleparallel scalar-tensor theory. The latter cosmological solution
possesses two main epochs for the universe: a scaling solution, which can
describe the matter or the radiation era, and an accelerated solution
described by a de Sitter universe as a future attractor.

In a future study, we plan to investigate further the physical properties of
these models. Moreover, the duality transformation will be applied to
investigate the physical properties of spacetime in the pre-big bang epoch
by using the late-time behaviour.

\begin{acknowledgments}
The author thanks the support of VRIDT through Resoluci\'{o}n VRIDT No.
096/2022 and Resoluci\'{o}n VRIDT No. 098/2022. Part of this study was
supported by FONDECYT 1240514.
\end{acknowledgments}

\end{document}